\documentclass[12pt,preprint]{aastex}
\shorttitle{CANGAROO PSR1259-63 Observations}
\shortauthors{Kawachi et al.}
\begin{document}
\title{A Search for TeV Gamma-ray Emission from 
the PSR B1259$-$63/SS2883 Binary System with the CANGAROO-II 10-m
Telescope} 
\author{A.~Kawachi\altaffilmark{1}, 
T.~Naito\altaffilmark{2}, 
J.R.~Patterson\altaffilmark{3}, 
P.G.~Edwards\altaffilmark{4}, 
A.~Asahara\altaffilmark{5}, 
G.V.~Bicknell\altaffilmark{6},
R.W.~Clay\altaffilmark{3}, 
R.~Enomoto\altaffilmark{1}, 
S.~Gunji\altaffilmark{7}, 
S.~Hara\altaffilmark{1,5}, 
T.~Hara\altaffilmark{2}, 
T.~Hattori\altaffilmark{8}, 
Sei.~Hayashi\altaffilmark{9},
Shin.~Hayashi\altaffilmark{9},
C.~Itoh\altaffilmark{10},
S.~Kabuki\altaffilmark{1}, 
F.~Kajino\altaffilmark{9},
H.~Katagiri\altaffilmark{1}, 
T.~Kifune\altaffilmark{11},
L.~Ksenofontov\altaffilmark{1}, 
H.~Kubo\altaffilmark{5},
J.~Kushida\altaffilmark{5,12},
Y.~Matsubara\altaffilmark{13},
Y.~Mizumoto\altaffilmark{14},
M.~Mori\altaffilmark{1},
H.~Moro\altaffilmark{8},
H.~Muraishi\altaffilmark{15},
Y.~Muraki\altaffilmark{13},
T.~Nakase\altaffilmark{8},
D.~Nishida\altaffilmark{5}, 
K.~Nishijima\altaffilmark{8}, 
M.~Ohishi\altaffilmark{1},
K.~Okumura\altaffilmark{1},
R.J.~Protheroe\altaffilmark{3}, 
K.~Sakurazawa\altaffilmark{12},
D.L.~Swaby\altaffilmark{3}, 
T.~Tanimori\altaffilmark{5}, 
F.~Tokanai\altaffilmark{7}, 
K.~Tsuchiya\altaffilmark{1}, 
H.~Tsunoo\altaffilmark{1}, 
T.~Uchida\altaffilmark{1}, 
A.~Watanabe\altaffilmark{7}, 
S.~Watanabe\altaffilmark{5}, 
S.~Yanagita\altaffilmark{10},
T.~Yoshida\altaffilmark{10},
and T.~Yoshikoshi\altaffilmark{16}
}
\affil{}
\email{kawachi@icrr.u-tokyo.ac.jp}
\altaffiltext{1}{
Institute for Cosmic Ray Research, University of Tokyo,  Kashiwa, 
 Chiba 277-8582, Japan}
\altaffiltext{2}{
Faculty of Management Information, Yamanashi Gakuin University, Kofu, Yamanashi 400-8575, Japan}
\altaffiltext{3}{
Department of Physics and Mathematical Physics, University of
Adelaide, SA 5005, Australia}
\altaffiltext{4}{
 Institute of Space and Astronautical Science, Sagamihara, Kanagawa 229-8510, Japan} 
\altaffiltext{5}{
Department of Physics, Graduate School of Science, 
Kyoto University, Sakyo-ku, Kyoto 606-8502, Japan}
\altaffiltext{6}{
Research School of Astronomy and Astrophysics, 
Australian National University, ACT 2611, Australia}
\altaffiltext{7}{
Department of Physics, Yamagata University, Yamagata, Yamagata 990-8560, Japan}
\altaffiltext{8}{
Department of Physics, Tokai University, Hiratsuka, Kanagawa 259-1292, Japan}
\altaffiltext{9}{
Department of Physics, Konan University, Kobe, Hyogo 658-8501, Japan}
\altaffiltext{10}{
Faculty of Science, Ibaraki University, Mito, Ibaraki 310-8512, Japan}
\altaffiltext{11}{
Faculty of Engineering, Shinshu University, Nagano, Nagano 480-8553, Japan} 
\altaffiltext{12}{
Department of Physics, Tokyo Institute of Technology, Meguro, Tokyo 152-8551, Japan}
\altaffiltext{13}{
Solar-Terrestrial Environment Laboratory,  Nagoya University, Nagoya, Aichi 464-8602, Japan} 
\altaffiltext{14}{
National Astronomical Observatory of Japan, Mitaka, Tokyo 181-8588, Japan}
\altaffiltext{15}{
Ibaraki Prefectural University of Health Sciences, Ami, Ibaraki 300-0394, Japan} 
\altaffiltext{16}{
Department of Physics, Osaka City University, Osaka, Osaka 558-8585, Japan}

\begin{abstract}
 Observations of the PSR\,B1259$-$63/SS2883 binary system 
 using the CANGAROO-II Cherenkov telescope are reported. 
 This nearby binary consists of a 48\,msec radio pulsar 
 in a highly eccentric orbit around a Be star, 
 and offers a unique laboratory to investigate
 the interactions between the outflows of the pulsar and Be star 
 at various distances.
 It has been  pointed out that the relativistic pulsar wind and 
 the dense mass outflow of the Be star may result in the emission
 of gamma rays up to TeV energies.
 We have observed the binary in 2000
 and 2001, $\sim$47 and $\sim$157 days after the October 2000 periastron.
 Upper limits at the 0.13--0.54 Crab level are obtained.
 A new model calculation for high-energy gamma-ray emission
 from the Be star outflow is introduced 
 and the estimated gamma-ray flux considering Bremsstrahlung, 
 inverse Compton scattering, and 
 the decay of neutral pions produced in proton-proton interactions, 
 is found to be comparable  to the upper limits of these observations. 
 Comparing our results with these model calculations, 
 the mass-outflow parameters of the Be star are constrained.
 \end{abstract}

\keywords{binaries: close---gamma rays: observations, theory---
stars: winds, outflows}

\section{Introduction}
\label{sect:intro}
 PSR B1259$-$63 (($\alpha,\delta$)(J2000) $=$ (13$^h$02$^m$47$^s$.68, 
 $-$63\arcdeg50\arcmin08\arcsec.6))
 is a 48\,msec radio pulsar 
 discovered in a 1500\,MHz radio survey of the southern Galactic
 plane\,\citep{johnston92a} which was subsequently found to be in 
 a highly eccentric orbit with a 10th magnitude main-sequence 
 star, SS\,2883\,\citep{johnston92b,johnston94}. 
 With an orbital eccentricity of 0.87,  
 the separation of the stars varies in the range
 0.97$\sim$14.0$\times {\rm 10}^{13}$cm 
 during the orbital period of 1236.72 days. 
 The periastron epoch is MJD~48124.35\,\citep{wex98}.
 SS\,2883 is of spectral type B2e\,\citep{johnston94}, 
 with a mass $M_*$ of $\sim$10\,$M_{\sun}$ and a radius $R_*$ 
 of $\sim$6\,$R_{\sun}$. 
 The luminosity and radius of the 
 B2e star correspond to an effective temperature $T_{\rm eff}$ 
 of $\sim$27,000~K at the star surface\,\citep{tavani97}.
 Its characteristic emission disc extends to at least 20\,$R_*$, 
 similar to the distance between the pulsar and the Be star at 
 periastron.  
 Here we assume a distance of 1.5\,kpc to the binary system, which has been 
 estimated from optical photometric observations of 
 SS\,2883\,\citep{johnston94}.

 The periastron passages have been closely observed at radio frequencies 
 \,\citep{johnston99,connors02}. 
 No pulsed emission was detected for about five weeks  
 centered on periastron and the pulsed emission was
 depolarized for $\sim$200 days also centered on 
 periastron. Timing measurements have shown that the disc of 
 the Be star is likely to be inclined with respect 
 to the orbital plane \,\citep{melatos95,wex98}, 
 which has been suggested in \citet{kaspi95,tavani97}.
 In \citet{connors02}, 
 the unpulsed light curves are discussed with an assumption 
 of two short-time crossings of the pulsar and the disc, 
 before [($\tau - $18~d) $\sim$ ($\tau - $8~d)] and after [($\tau +$12~d) 
 $\sim$ ($\tau +$22~d)]  periastron ($\tau$).

 A weak X-ray signal was first detected by 
 {\it ROSAT} which observed the system just after apastron in
 September 1992\,\citep{cominsky94}.
 Through 1994--1996, 
 unpulsed X-ray emission with a single power-law spectrum was
 detected 
 at the six different orbital phases observed by  {\it ASCA}
 \,\citep{hirayama96,hirayama99}. 
 The photon index of the X-ray spectrum 
 is about $-$1.6 in the post-periastron to apastron period, 
 steepening towards periastron 
 where the steepest index of $-$1.96 was observed.
 The 1--10~keV band luminosity varies by about an order of 
 magnitude, from $\sim$10$^{34}$ ergs s$^{-1}$ around periastron to 
 $\sim$10$^{33}$ ergs s$^{-1}$ at apastron. The maximum luminosity 
 was detected at $\tau -$12~d, with the intensity decreasing
 at periastron, then increasing again. 
 The column density was low and constant 
 ($\rm{6}\times\rm{10}^{21}\rm{cm}^{-2}$) at all orbital phases.
 The periastron passage in January 1994 was monitored
 by a multi-wavelength 
 campaign including observations in the X-ray and gamma-ray bands 
 with {\it ROSAT}, {\it ASCA} and {\it CGRO} \citep{grove95}.
 The power-law spectrum (photon index $\sim -$2.0) extended 
 to the 200\,keV energy region of {\it OSSE},
 with no pulsations being detected.  
 No emission in the energy range of 1\,MeV--3\,GeV was detected 
 down to the observational limits. 
 {\it OSSE} failed to detect signals at the apastron passage in 1996, however, 
 its upper limit does not conflict with the extrapolation of the {\it ASCA}
 spectrum\,\citep{hirayama99}.
 Several TeV observations of the binary system were performed 
 in 1994 and 1997 using 
 the CANGAROO 3.8-m ground-based Cherenkov telescope, 
 resulting in a marginally significant  
 suggestion of gamma-ray signals \citep{sako97}.

 The multi-wavelength spectrum from the 1994 periastron strongly 
 implies that the 
 hard X-ray emission up to 200\,keV originates from synchrotron radiation 
 of non-thermal electrons\,\citep{tavanikaspi94}. Electrons 
 released in the pulsar wind may be accelerated in a shock wave  
 generated in the region where the relativistic pulsar wind interacts 
 with dense mass flow from the Be star.
 Adjusting the pressure balance between the flows,  
 \citet{tavani97} have interpreted the measured hard X-ray spectrum on the 
 basis of 
 accelerated particles in the pulsar-side shock, 
 using an approximated approach to the Klein-Nishina effect for emission.
 They conclude that the energy loss of electrons 
 due to inverse Compton scattering is dominant, 
 and that the Lorentz factor of accelerated electrons is
 $\Gamma_{\rm e} = {\rm 10}^6-{\rm 10}^7$.
 Accretion onto the neutron star is unlikely to be significant, as there is 
 an absence of X-ray/gamma-ray pulsations, 
 an absence of the day--scale fluctuations in X-rays, 
 a relatively low X-ray luminosity,  
 and negligible absorption as a result of the low column density.
 The consistency of the X-ray luminosities at the same orbital phases
 in different years supports the idea that the observed time
 variability is due to binary modulation \citep{hirayama99,kaspi97}.
 Recently, another model for the pulsar-side shock 
 considering the Klein-Nishina effect in the emission and cooling process 
 has been proposed. \citet{shibazaki02} note that 
 the inverse Compton cooling dominated spectrum is flatter, 
 since the Klein-Nishina effect suppresses the cooling of 
 higher-energy electrons.
 It is argued that synchrotron cooling, instead of the
 inverse Compton scattering discussed by \citet{tavani97}, 
 is the dominant process for the energy loss of electrons 
 in the pulsar wind 
 on the account of the steepening of the X-ray spectral index observed 
 around the 1994 periastron.  

 The light curves of the radio unpulsed emission around periastron 
 have been recently modeled by the adiabatic expansion of synchrotron 
 bubbles formed in 
 the pulsar and the Be star disc interaction \citep{connors02}. 
 They assume short-time interactions of the pulsar and the disc, as 
 the pulsar should cross the disc twice in the orbital period.
 When the pulsar enters the disc, electrons are accelerated 
 in the contact surface of the pulsar wind and the disc material,
 but after the pulsar leaves the disc, the pulsar-wind bubble remains  
 behind, moves in the disc-flow, and decays through synchrotron losses.
 The model successfully explains the radio data, however, 
 it does not appear to describe the X-ray data well;
 for example, the weak unpulsed 
 emissions in the X-ray region was observed after these modeled bubbles 
 should have decayed by adiabatic expansion as moving outwards,
 and, the constant spectral index in the radio region is inconsistent with 
 the steepening observed in the X-ray spectrum.

 Electrons accelerated in a pulsar-side shock to 
 Lorentz factors of $\Gamma_{\rm e} ~\gtrsim \rm{10}^6$ 
 in the radiative environment of the binary system may produce 
 high energy gamma rays.
  The suggestion that detectable levels of gamma-ray emission may
  arise in the shocked pulsar wind via inverse Compton scattering
  \citep{kirkball99} provided the initial motivation for the
  observations described here. Subsequently, inverse Compton emission from
  the un-shocked region of the
  pulsar wind has been considered \citep{ballkirk00,balldodd01}.
  The integrated contribution 
  from the un-shocked pulsar wind may increase the gamma-ray flux 
  around periastron for some conditions. The maximum level of 
 emission in the TeV energy range
  is estimated to be 
  $\sim{\rm 4} \times {\rm 10}^{-5}{\rm MeV cm}^{-2}{\rm s}^{-1}$ 
 in the integrated energy flux with a wind Lorentz 
  factor of 10$^7$\,\citep{ballkirk00} which may raise the TeV
  gamma-ray flux above our detector's sensitivity of typically
  ${\rm 10}^{-11}-{\rm 10}^{-12}{\rm TeV cm}^{-2}{\rm s}^{-1}$.
  Further studies have considered the effect of the termination of
  the wind \citep{balldodd01}, which, depending on its assumped
  location, may act to decrease the inverse Compton flux compared
  to previous predictions.
 On the other hand, 
 \citet{shibazaki02} predict a maximum energy flux from the pulsar wind of  
 $\sim {\rm 10}^{-13} {\rm erg \,s}^{-1}{\rm cm}^{-2}$  
 at TeV energies around periastron, which would require a very 
 deep observation to detect.

 At the contact surface of the pulsar and Be star flows,  
 ions and electrons in the Be star outflow
 may be accelerated to high energies 
 via the first-order Fermi mechanism. 
 In the dense outflow from the Be star there is a lot
 of target material for proton-proton interactions and 
 Bremsstrahlung emission.
 The Be star also provides
 target photons for upscattering by the inverse Compton mechanism, 
 in addition to the 2.7\,K microwave background radiation.
 The densities of these targets increase 
 as the contact surface gets closer to the Be star, and so does the 
 total energy of accelerated particles. 
 In this paper, a new model calculation for gamma-ray emission
 from the accelerated particles in the Be star outflow, taking into 
 consideration Bremsstrahlung, the inverse 
 Compton mechanism, and proton-proton interactions, is applied to 
 the binary system and discussed along with our observational results. 

\section{Observations}
 The CANGAROO 
 ($\underline{C}$ollaboration between $\underline{A}$ustralia and
 $\underline{N}$ippon for a $\underline{GA}$mma-$\underline{R}$ay
 $\underline{O}$bservatory in the $\underline{O}$utback) Cherenkov 
 telescopes are located near Woomera, South Australia 
 (136\arcdeg 47\arcmin E, 31\arcdeg 06\arcmin S, 160m a.s.l.).
 The 3.8-m CANGAROO telescope, 
 used for the previous observations of PSR\,B1259$-$63, 
 was operated from 1991 to 1998.  
 The CANGAROO-II telescope 
 was constructed in 1999 initially with a 7-m diameter dish, 
 which was upgraded to 10-m in 2000\,\citep{mori01}.
 Cherenkov photons from extensive air showers which 
 are initiated by primary gamma-rays/cosmic rays 
 are collected with a parabolic  
 reflector and detected by an imaging camera placed in the prime focal 
 plane.  The 10-m reflector 
 consists of 114 spherical plastic mirrors of 0.8\,m 
 diameter\,\citep{kawachi01} to make a composite parabolic shape ($f$/0.8).
 The camera has an
 array of 552 photomultiplier tubes (PMTs) 
 of half-inch
 diameter (Hamamatsu, R4124UV) 
 covering a field-of-view of 2.8\arcdeg,
 with groups of 16 PMTs using the same high voltage line. 
 In the observations described in this paper, 
 accepted events were required to meet two conditions 
 within the central region of the camera (1$\arcdeg$ diameter); 
 at least three PMTs 
 above a $\sim$3-photoelectron threshold, and at least 
 one preamplifier unit with an analogue sum of signals 
 above a $\sim$10-photoelectron threshold. 
 The typical raw trigger rates of these observations were in the 10--80~Hz
 range.
 The charge and timing information of each PMT signal was recorded 
 for each event. 
 Details of the telescope are given in \citet{mori01} and \citet{tanimori_NIM}.

\begin{figure}
\plotone{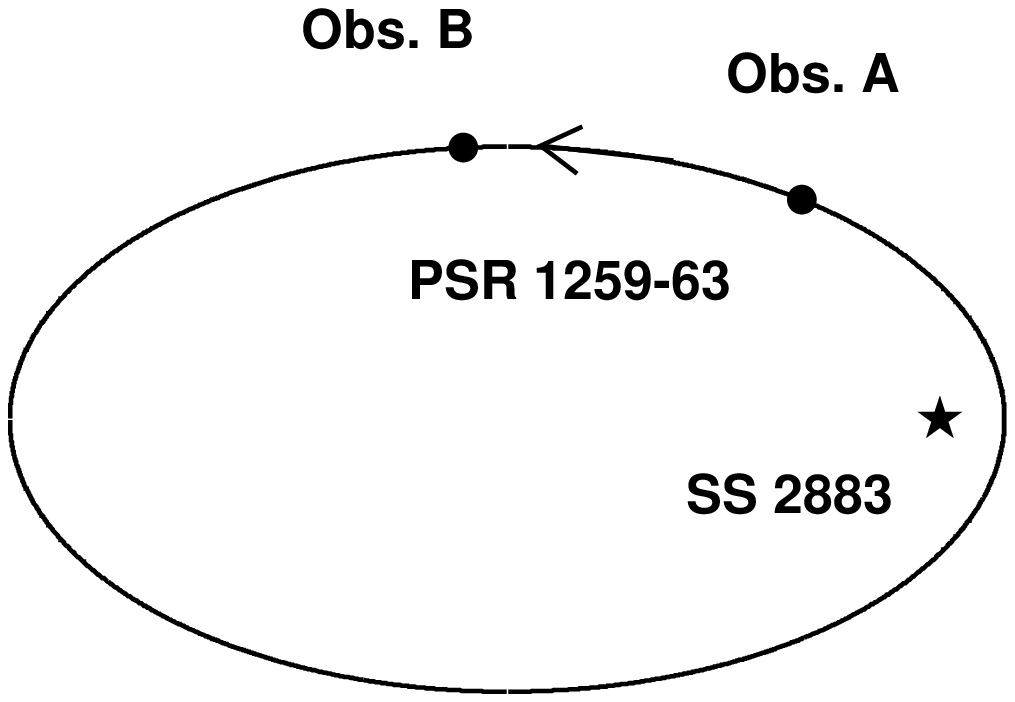}
\caption{
 Approximate location of the pulsar at 
 the observed orbital phases for $Obs.\,A$ (December 2000) and 
 $Obs.\,B$ (March 2001) are marked as filled circles on the schematic orbit. 
\label{fig:obs_orbit}
}
\end{figure}

 The PSR\,B1259$-$63/SS2883 system was observed with the 
 CANGAROO-II Cherenkov telescopes at two different orbital phases;
 for several days in December 2000 (hereafter $Obs.\,A$;
 MJD 51881.4 in average of the observation time) 
 and in March 2001 ($Obs.\,B$; MJD 51991.5), 
 about 47 and 157 days after the periastron of October 
 2000 (MJD 51834.51). The orbital phase of $Obs.\,A$ is similar to 
 the {\it ASCA} Obs.\,4 in \citet{hirayama99}. 
 We have no data closer to periastron as
 observing conditions were not suitable 
 during June--November 2000.
 The orbital phases of the observations are schematically shown in 
 Figure\,\ref{fig:obs_orbit}, and details of the
 observations are summarized in Table\,\ref{table:obs_summary}. 
 In all observations the telescope was pointed so that the binary system 
 was at the tracking center.

 The target and offset region(s) were observed for equal amounts 
 of time on each night under moonless and fine sky conditions.
 In a typical observation, the target (ON-source) region was observed
 for a few hours including source culmination,  
 and an offset (OFF-source) run was carried out before and/or after the 
 ON-source run to cover the same track, in the elevation and 
 azimuthal angles, as that of the ON-source run.
 The average zenith angles were 
 58$\arcdeg$ for $Obs.\,A$ and 34$\arcdeg$ for $Obs.\,B$, 
 corresponding to  differences in observing seasons.
 The observations closer to periastron, 
 $Obs.\,A$, were performed at larger zenith angles, thus at a higher energy
 threshold, than the optimum observing condition for the source.
 The zenith angle of 58$\arcdeg$ is similar to the 
 observing conditions for the Crab nebula 
 ($\sim$55$\arcdeg$ in \citet{tanimori_crab}) from the CANGAROO site, 
 thus the data has been analyzed in a similar manner to the recent analysis 
 of the Crab data \citep{itoh03}.

\section{Analysis and Results}
\subsection{Data Analysis}
 The digitized counts of PMT signal charges have been calibrated after 
 pedestal subtraction.   The gains of the pixels have been normalized 
 using a blue LED located 
 at the center of the telescope to illuminate the camera uniformly. 
 The LED is driven by a fast pulse generator.
 Accidental events, caused mainly due to the night sky or environmental 
 background light, have been removed in the analysis by requiring 
 at least five neighboring PMTs 
 to exceed the  $\sim$3.3 photoelectron threshold 
 within $\pm$35\,nsec. 
 A small number of pixels affected by noise or the passage of bright
 stars have been excluded from the analysis.
 The count rate of events which satisfy
 these criteria is stable (about 0.6\,Hz and 1.9\,Hz in 
 $Obs.\,A$ and $Obs.\,B$ data sets, respectively), and differences
 in the raw event rates between ON- and OFF-source runs, ratios of
 about 1:4 in $Obs.\,A$ and 3:2 in $Obs.\,B$, respectively, 
 have been resolved. 
 Data under cloudy or unstable conditions have been rejected 
 by checking deviations of the counting rate.
 The effective observation time and event numbers 
 at this stage of the analysis are summarized 
 in Table\,\ref{table:data_summary}. 

 In order to discriminate gamma-ray events from cosmic-ray 
 induced events, a likelihood analysis has been 
 applied \citep{enomoto_sim2,itoh03} to the characteristic 
 light images recorded by the camera.
 In this image-analysis, we also require
 the total charge contained in an image to be greater than 
 $\sim$35~photoelectrons, in order to improve the 
 efficiency of background rejection.
 We have used Monte Carlo simulations \citep{enomoto_sim1}
 for a gamma-ray ($\gamma$) event set  and 
 the OFF-source events for the background  ($BG$) event set.
 The average zenith angle, the discarded pixels, and 
 the cut parameters in the analysis have been taken into account 
 in the simulations for the different conditions of 
 $Obs.\,A$ and $Obs.\,B$.
 The following results are based on simulations assuming 
 a power-law energy spectrum with an index of $-$2.5 for the generated 
 gamma-rays. 
 The ``hit-map'' of triggered pixels, weighted by size of the signal, 
 has been fitted with an ellipse parametrized by
 the r.m.s.\ spread of light along the minor/major axis of the image 
 ($width/length$), 
 and the distance between the image centroid and the source position
 ($distance$) \,\citep{hillas82} to characterize each event.
 Images truncated by the camera edge, or too concentrated 
 at the camera center, have been omitted by putting a loose limit
 on  $distance$ (0.35\arcdeg $\le distance \le$ 1.2\arcdeg).
 Each of the image parameters have been 
 plotted against
 the total signal in the image 
 for the $\gamma$ and $BG$ event sets to produce 
 probability density functions (PDF) of the ``$\gamma$-like'' 
 and ``$BG$-like'' events, respectively.
 Thus the dependences of the image parameters 
 on the event size have been taken into the analysis. 
 Finally, a single parameter, 
 $R_{prob} = Prob(\gamma)/[Prob(\gamma)+Prob(BG)]$, 
 is calculated for each event,
 where $Prob(\gamma)$[$Prob(BG)$] is the probability of the event
 being due to a $\gamma$ [$BG$], calculated from the two-dimensional PDFs.
 We use the PDFs of three parameters with equal weight.
 A selection criteria ($R =$0.4; $R_{prob} \ge R$) has been chosen 
 considering the acceptance of $\gamma$ events and 
 the figure-of-merit of $\gamma$ events to the $BG$ events. 
 The event numbers of the selected data are listed in 
 Table\,\ref{table:data_summary}.

\begin{figure}[ht]
\plottwo{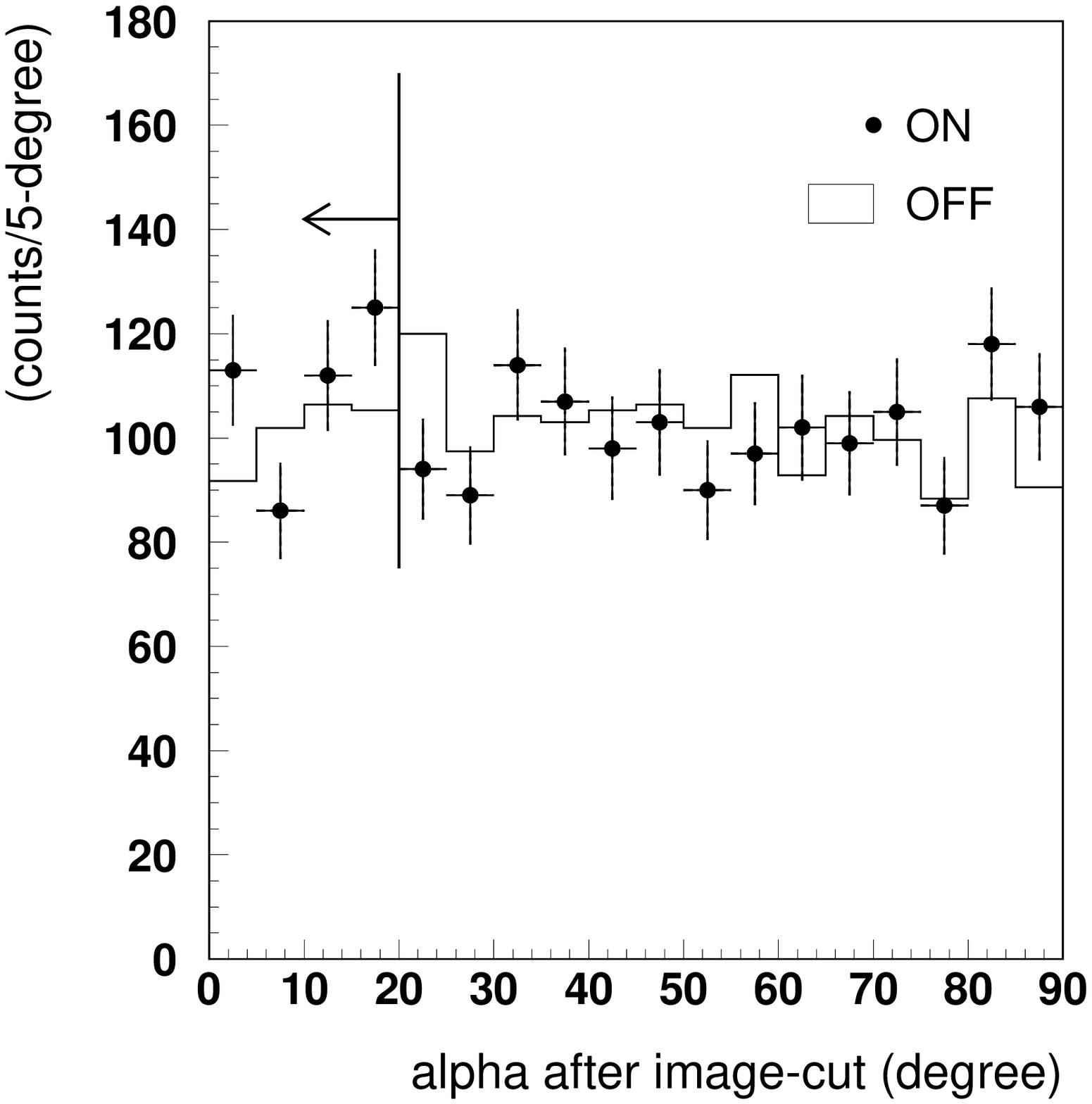}{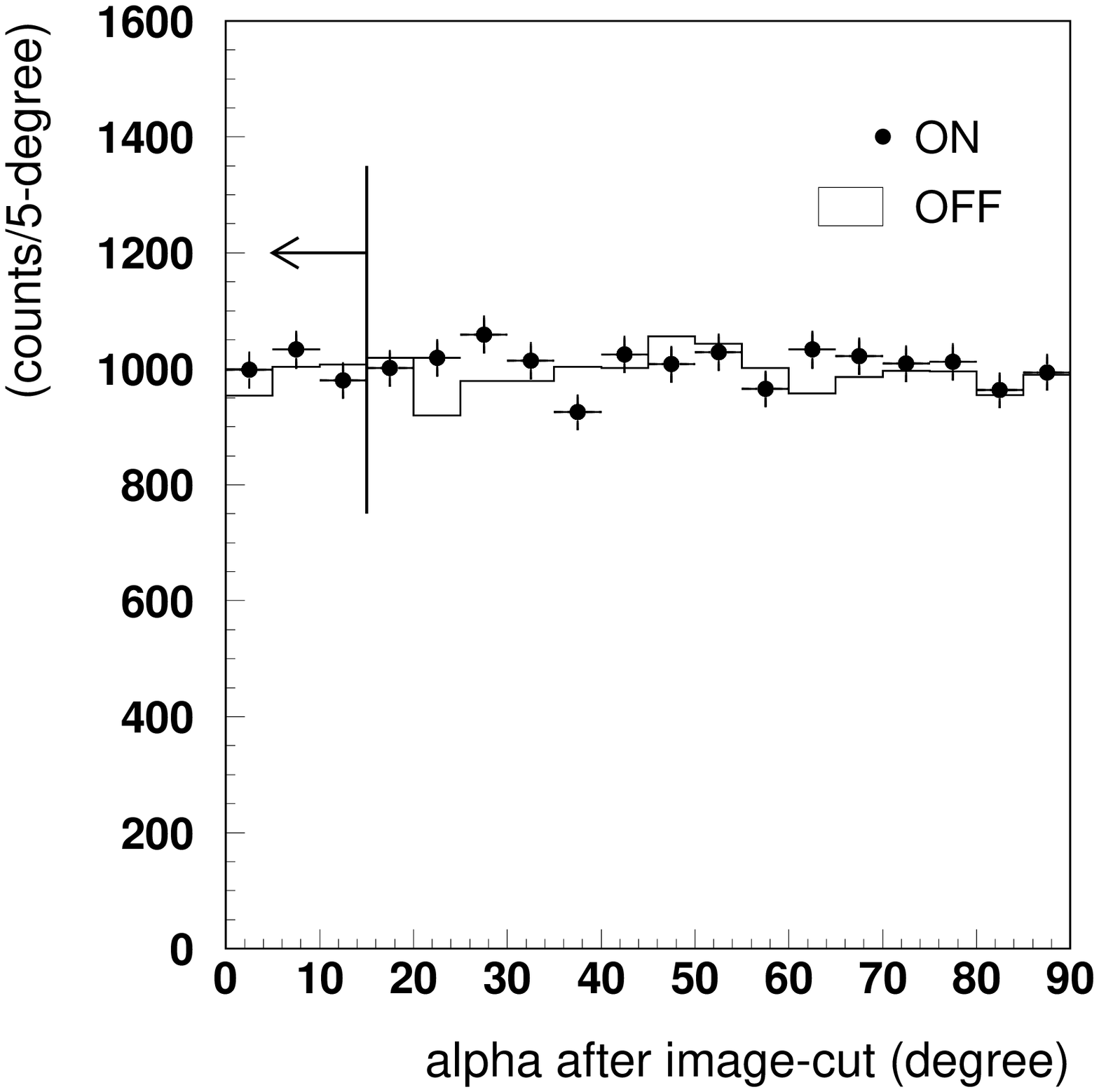}
\caption{Distributions of the orientation angle $alpha$. 
 ``OFF" (blank histogram) and ``ON" (filled circles 
 with statistical errors) are plotted after normalization. 
 The region $\alpha \leq \alpha_{cr}$ is assumed to contain 
 gamma-ray signals from the binary.  
 $(left):$ $Obs.\,A$ data set. $\alpha_{cr} = {\rm 20}\arcdeg$. 
 $(right):$ $Obs.\,B$ data set. $\alpha_{cr} = {\rm 15}\arcdeg$. 
\label{fig:alpha_distribution}
}
\end{figure}

 After the image selection, gamma-rays from the pulsar should have 
 the image orientation parameter 
 $alpha$ \citep{punch92} less than 20$\arcdeg$ 
 for the $Obs.\,A$ data set, and 15$\arcdeg$ for the $Obs.\,B$ set. 
 A broader $alpha$ distribution for gamma-rays at 
 larger zenith angle is expected from simulations\,\citep{okumura_mrk421}.
 Figure\,\ref{fig:alpha_distribution}
 shows the distributions of 
 $alpha$ after all the other cuts have been applied.
 The OFF-source (``OFF") $alpha$ distributions
 are normalized to that of the ON-source (``ON") by the number of 
 events with  $alpha \ge 40\arcdeg$. 
 The normalization factor is consistent with that deduced 
 from the effective observation times, within statistical errors. 
 No statistically significant excess of the ``ON"  over 
 the ``OFF" is seen in either of the $alpha$ histograms.
 Subtracting the normalized ``OFF" counts
 from the ``ON" data results in 
 31 and 47 events within the $alpha$ selection criteria, 
 corresponding to the significances 
 (assuming Poisson fluctuations only) of
 $+$1.0$\sigma$ and $+$0.60$\sigma$, for $Obs.\,A$ and
 $Obs.\,B$ respectively.

\subsection{Energy Threshold and Upper Limit Flux}
\begin{figure}[ht]
  \plotone{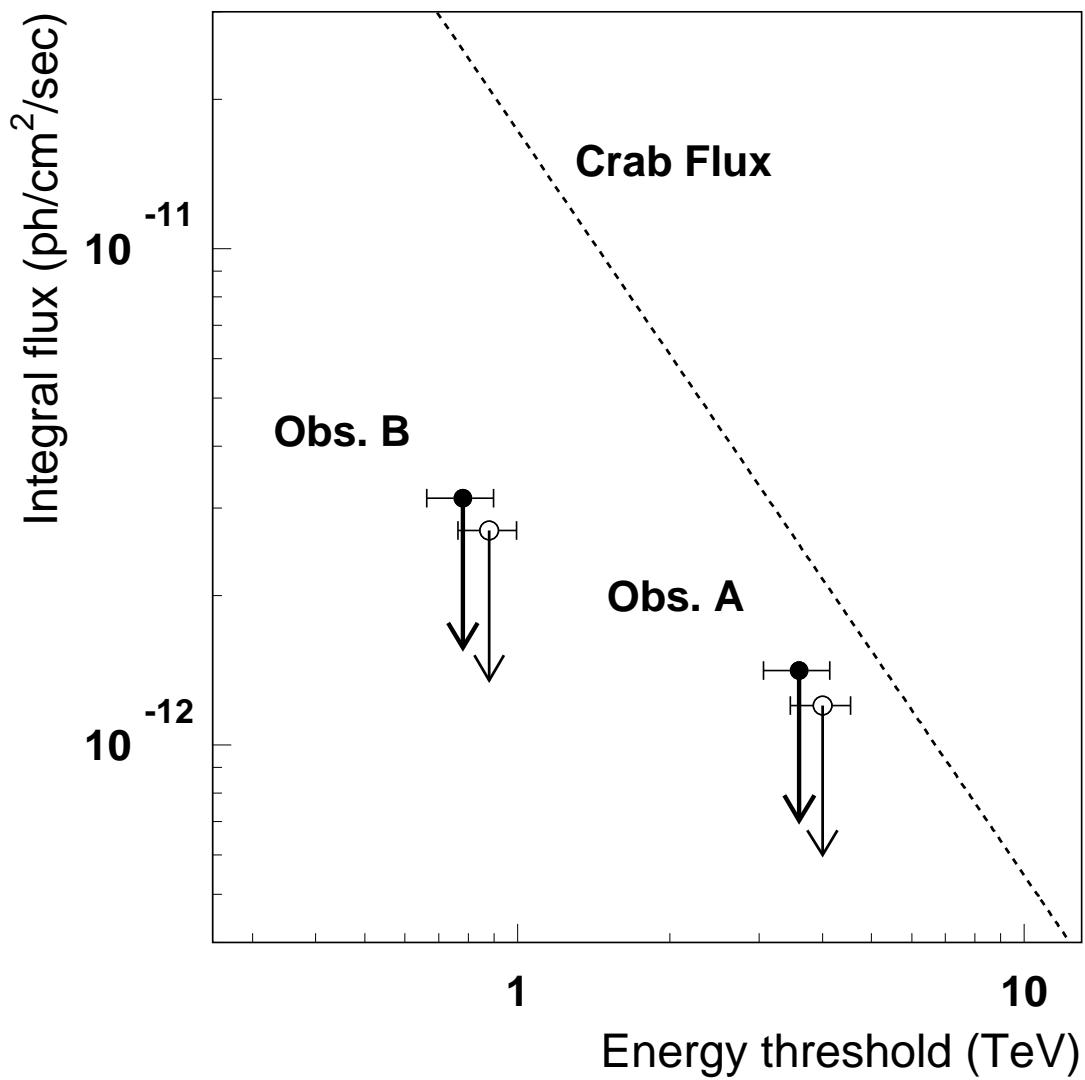}
\caption{Obtained 2-$\sigma$ upper limits of the 
 integral flux from  CANGAROO-II observations.
 The two observation data sets ($Obs.\,A$ and 
 $Obs.\,B$) have different energy thresholds corresponding to the 
 different zenith angles of the observations. 
 The results obtained assuming a 
 power-law spectral index of $-$2.5 in the Monte Carlo simulations, 
 are plotted in closed circles, 
 and these with a spectral index of $-$2.0 are shown in open circles.
 The integral flux  of the Crab nebula, calculated from the 
 differential flux \protect{\citep{tanimori_crab}} is 
 plotted for comparison. 
\label{fig:integral_flux}
}  
\end{figure}

 The gamma-ray acceptance for the cuts used in the analysis has been 
 estimated  based on the simulations.
 The energy threshold is defined as the peak of the acceptance 
 multiplied by the generated energy spectrum, and thresholds of 
 3.6\,TeV and 0.78\,TeV are derived for a $E^{-2.5}$ spectrum for 
 the $Obs.\,A$ and $Obs.\,B$ data sets, respectively. 
 The corresponding effective areas are 3.6$\times$10$^9$~cm$^{2}$ 
 and 1.3$\times$10$^9$~cm$^{2}$.
 The 2$\sigma$ upper limits of our result are 
 \begin{quote}
$F(\geq 3.6TeV) \leq 1.4\times10^{-12}{\rm cm}^{-2}{\rm s}^{-1}$~~($Obs.\,A$)
\end{quote}
and 
\begin{quote}
$F(\geq 0.78TeV) \leq 3.1\times10^{-12}{\rm cm}^{-2}{\rm s}^{-1}$~~($Obs.\,B$).
\end{quote}
 Figure\,\ref{fig:integral_flux} shows  
 the upper limits of the two observations.
 The integral flux of the Crab nebula is also shown 
 \citep{tanimori_crab} as a reference. 
 The systematic error in the energy scale 
 determination has been estimated to be 15 percent\,\citep{itoh03} 
 and is shown in the figure as errors in the abscissa.

The whole procedure of the analysis has been performed changing 
 the simulated spectral index from  $-$2.5 to $-$2.0. 
The significance levels of the gamma-ray signals are unchanged, 
and the corresponding threshold energies and the gamma-ray acceptances  
increase by about 20 percent. Thus the 
 upper limits assuming a spectral index of $-$2.0 are;  
$F(\geq 4.0TeV) \leq 1.2\times10^{-12}{\rm cm}^{-2}{\rm s}^{-1}$~($Obs.\,A$)
 and 
$F(\geq 0.88TeV) \leq 2.7\times10^{-12}{\rm cm}^{-2}{\rm s}^{-1}$~($Obs.\,B$), 
 respectively, as shown in Figure\,\ref{fig:integral_flux}.

\section{Discussion}
 The observational results are compared with some model calculations. 
 A new model of gamma-ray emissivity is introduced, 
 considering the particles accelerated in the Be star outflow.

\subsection{Models of the two flows}  
\begin{figure}[ht]
\plotone{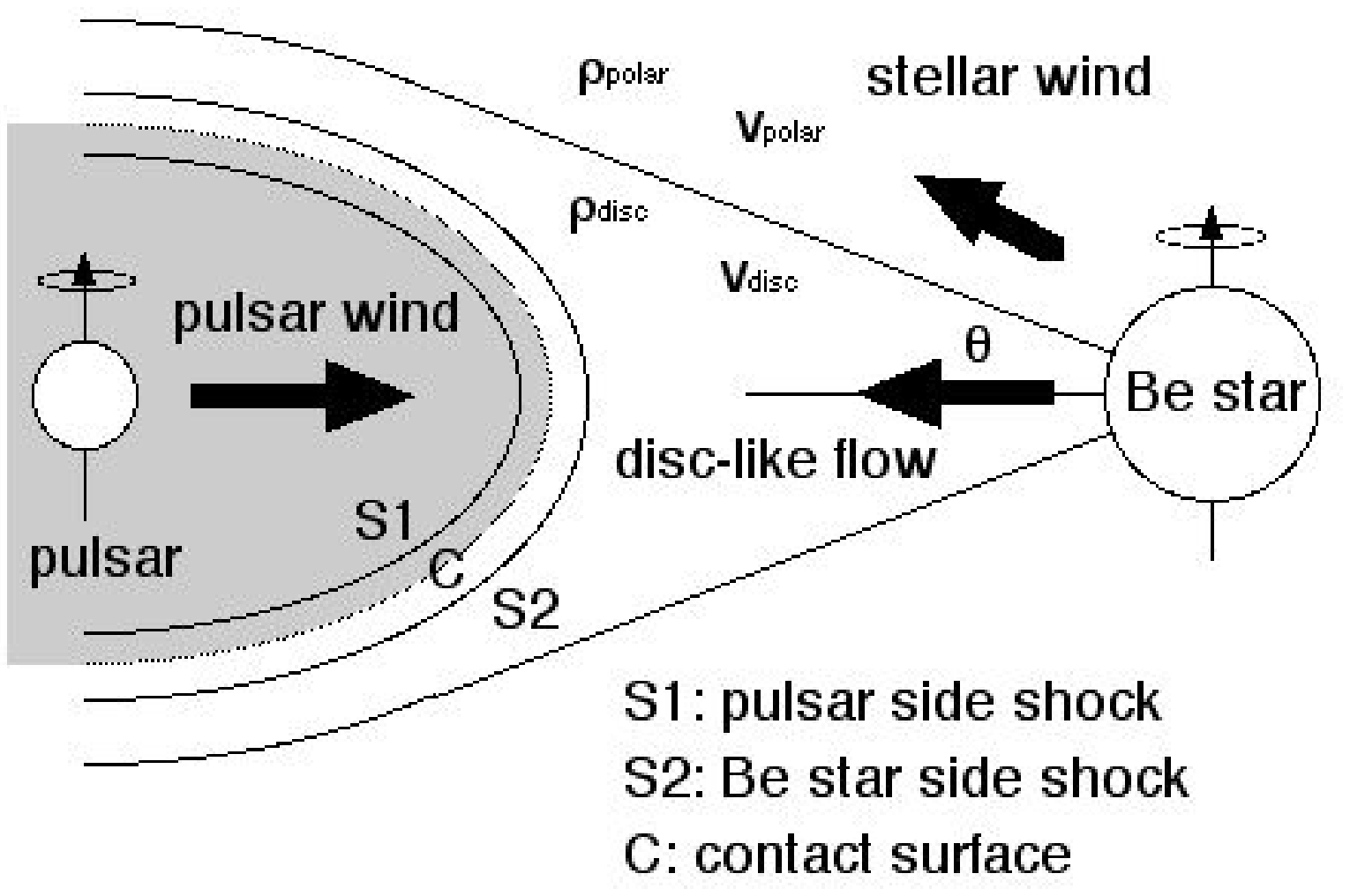}
\caption{Schematic image of the binary system. The wind from the
 pulsar and the disc-like outflow from the Be star 
 results in a region of pressure balance between the stars. 
\label{fig:binary_system_image}
}
\end{figure}

 Figure\,\ref{fig:binary_system_image} schematically illustrates 
 the assumed configuration of the system: 
 the pulsar and its relativistic pulsar wind, 
 the Be star and its polar and disc-like outflows, 
 and the shock composed of three surfaces: 
 pulsar-side shock, contact surface, and Be-star-side shock.
 Particles are assumed to be 
 accelerated by the shock at the pressure balance between 
 the flows of the two stars.
 In the figure, the contact discontinuity 
 between the pulsar wind and the equatorial disc of the Be star 
 is illustrated.
  The alignment of the Be star disc to the orbital plane, and its effect,  
 will be discussed later in calculating light curves over 
 orbital phase.

 For the pulsar wind, we adopt the model of \citet{kennel84} for the
 synchrotron nebula around the Crab pulsar.  For the Be
 star mass-flow, the simple model of \citet{waters86} is used, 
 which represents radiations from Be stars using the IR, optical, 
 and UV observational results. 
 The parameters are chosen so as to be consistent with the observational 
 results of Be stars in general \citep{cote87} and of the 
 PSR B1259$-$63/SS2883 binary \citep{johnston94, johnston96, 
 melatos95}.
 We fully consider the Klein-Nishina effect in the calculations of 
 the emission processes via electrons.
 Provided the pulsar wind is driven by the 
 spin down luminosity ($\dot{E}_{\rm rot}$) of the pulsar,
 a fraction ($f_{\rm pw} =$  0.1) of  
 the wind luminosity is assumed to be enhanced in the equatorial plane.
 Both kinetic and electromagnetic energies are included in 
 $\dot{E}_{\rm rot}$ \,\citep{kennel84}.
 The radial distribution of 
 the wind pressure, $P_{\rm pw}$, is given by 
\begin{equation}
\label{eq:p_pw}
 P_{\rm pw}(r) = \frac{\dot{E}_{\rm rot}}{f_{\rm pw}4\pi r^2 c}~~, 
\end{equation}
 where $r$ is the distance from the pulsar and $c$ is
 the speed of light.

 For the mass-flow of the Be star, we consider a high-density, slow, 
 equatorially orbiting disc-like flow \citep{waters86}, and  
 a low-density, fast, polar component (stellar wind)
 \,\citep{waters88,dougherty94} as well.
 The density profile, $\rho$, is assumed to depend on the distance from 
 the center of the Be star, 
 $R$, as $\rho(R) = \rho_0 (R/R_*)^{-n}$ with a power-law 
 index $n$, where $R_*$ is the star radius and 
 $\rho_0$ is the density of the outflow at the surface of the star.  
 The flow speed $v(R) = v_0(R/R_*)^{n-2}$ is obtained 
 from conservation of mass flux, where $v_0$ is speed of the outflow 
 at the surface of the star. 
 Then the momentum flux of the flow, $P_{\rm Be}$, is 
\begin{equation}
 \label{eq:p_be}
 P_{\rm Be}(R) = \rho v^2 = \rho_0 v_0^2 (\frac{R}{R_*})^{n-4}.
\end{equation}
 In our calculation, indices $n$ of 2.5 and 2 
 are chosen in outflows of disc and polar wind, respectively.

 The location of the shock regime is determined by the balance 
 between pressures of the pulsar wind (Eq.\,\ref{eq:p_pw}) and 
 of the Be star outflow (Eq.\,\ref{eq:p_be}). 

 We introduce a new parameter, $x$, 
 defined as 
\begin{equation}
 x = \frac{\rho_0}{{\rm 10}^{-12} {\rm g cm}^{-3}}\frac{v_0}
 {{\rm 10}^6 {\rm cm s}^{-1}}. 
\end{equation}
 When $x$ is larger, the location of the pressure balance
 becomes further from the Be star.
 If we assume that the opening angle $\theta_{\rm disc}$
 of the disc outflow is 15\arcdeg\,\citep{johnston96}, 
 the parameter $x$ is related to the $\Upsilon$ 
 of \citet{tavani97} by $\Upsilon
 \equiv (\dot{M}/{\rm 10}^{-8}{M}_{\sun}\ {\rm yr}^{-1})\ 
 (v_{0}/{\rm 10}^6 {\rm cm\ s}^{-1}) 
 = {\rm 0.90} \times x
 \times (v_{0}/{\rm 10}^6 {\rm cm\ s}^{-1})$, 
 where $\dot{M}$ is the mass loss rate.
 The parameter $x$ depends on $v_0$ and $\rho_0$,
 which are obtained directly from UV/optical observations, 
 independent of the disc opening angle.
 As shown later, the gamma-ray emission is approximately 
 proportional to $x^2$ in our model.

\subsection{Particle acceleration and gamma-ray spectrum}
\label{sect:discussion-2}
 First, we deduce the flux $j_{\rm i}$ (i $=$ e,p)
 as a function of energy $E_{\rm i}$  of the particles
 in the Be star outflow on the basis of Fermi acceleration,
 where $e$ denotes electrons and $p$ denotes protons.
 For simplicity, we assume that all ions are protons.
 Secondly, we calculate the energy flux 
 from the binary system induced from
 the emission mechanisms of Bremsstrahlung, inverse Compton (electrons) and 
 proton-proton collisions. 

 In general terms, 
 the momentum spectrum  $dN/dP$ of the shock-accelerated particles 
 is expressed as $N_{0} P^{-\alpha}$ and 
 4$\pi j$ is obtained from  $v dN/dE_{\rm k}$, 
 where $E_{\rm k}$ is the kinetic energy, described as 
 $dN/dE_{\rm k} = dP/dE_{\rm k} dN/dP$.
 The constant $N_{\rm 0,i}$ is evaluated  with the following integration 
 regarding of the energy balance at the shock location in the Be star flow;
 \begin{equation}
 \label{eq:energy_balance-general}
   \int^{E^{\rm max_{\rm i}}}_{m_{\rm i} c^2}{\frac{dN_{\rm i}}{dE_{\rm i}}
   (E_{\rm i}) dE_{\rm i}} 
   =  f_{\rm acc, i} P_{\rm Be} (R_{\rm shock}), ({\rm i} = {\rm e,p}), 
 \end{equation}
 where $R_{\rm shock}$ is the distance to the 
 contact surface from the Be star center.
 We assume $f_{\rm acc, i} = $ 0.001
 and 0.1 is assumed for $i = e$ and $p$, respectively, as 
 the efficiency of the acceleration to be consistent with an e/p ratio 
 in cosmic ray observations (e.g. \citet{mullerproc,baring99}).
 The variation of $R_{\rm shock}$ causes the orbital modulation 
 in the light curve. 
 The orbital inclination to the line of sight has not been included in 
 our calculation, as this effect is less significant in 
 the Be star emission models than in the pulsar wind emission models.  
 Anisotropy of the optical photons from the Be star is neglected for 
 simplicity.
 Assuming Fermi acceleration, a power-law index of $\alpha = -$2.0
 is taken for the proton momentum spectrum $dN_{\rm p}/dP_{\rm p}$
 with an assumed compression ratio of 4.0. 
 The spectral index of the electron momentum spectrum 
 at the shock front does not 
 vary much from the canonical $\alpha = -$2.0 for plausible values of 
 pulsar wind parameters, because inverse Compton cooling
 in the higher energy electrons 
 are reduced by the Klein-Nishina effect \citep{shibazaki02}. 
 In addition, synchrotron cooling does not affect the spectral index,
 since the magnetic-field strength on the Be-star side should be weak.
 We therefore assume that the electron spectral index has a constant value of 
 $\alpha = -$2.0. 

\begin{figure}[ht]
\plotone{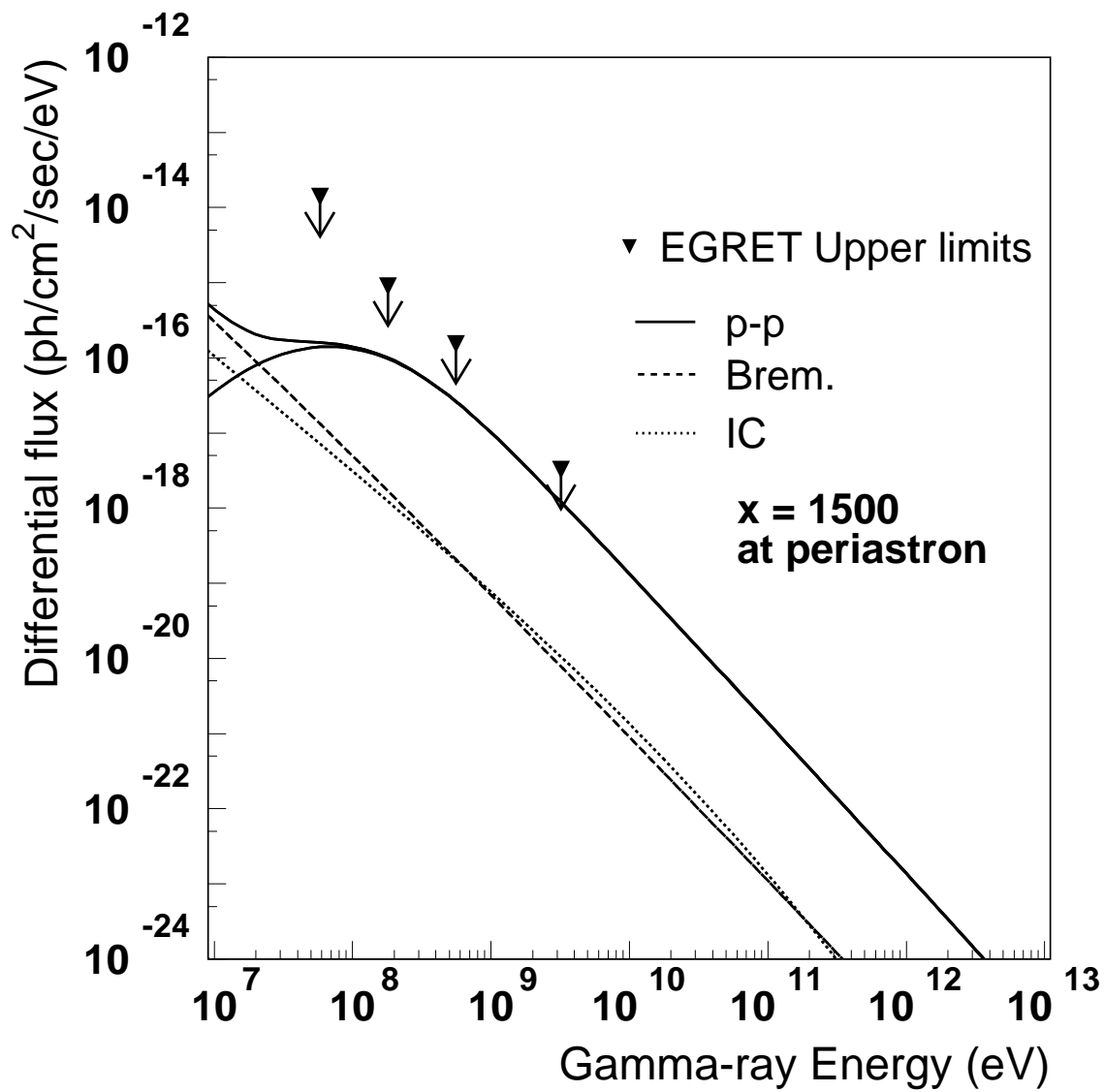}
 \caption{
 Differential flux of the periastron epoch is calculated 
 with $x_{\rm disc} =$ 1500, to be 
 compared with the upper limits by {\it EGRET} at the 
 periastron passage in January 1994 \protect\citep{tavani97}.
 The contributions from
 proton-proton collisions (p-p), Bremsstrahlung (Brem.),
 inverse Compton scattering (IC),  and the total flux are shown.
}  
\label{fig:spectrum_comparison}
\end{figure}

 The integration is performed from the threshold energy or 
 the particle mass, $m_{\rm e,p} c^2$, to 
 the maximum energy of the accelerated particle $E^{\rm max}_{\rm e,p}$, 
 which we assume here to be $\sim$10$^{15}$~eV.
 Applying the obtained $j_{\rm e,p}(E_{\rm e,p})$, 
 the gamma-ray spectrum from the 
 source at the distance $D$ is calculated. 
 For the proton-proton collision emission mechanisms, the spectrum is 
 calculated as  
\begin{equation}
\label{eq:gamma_flux-protons}
 F_{\gamma}^{\rm pp}(E_{\gamma}) = \frac{1}{D^2}\int{n_{\rm target} dV}
 \int\int{dE_\pi dE_{\rm p}} \frac{2}{p_\pi} j_{\rm p}(E_{\rm p})
\frac{d\sigma_{\rm pp\rightarrow\pi}(E_{\pi},E_{\rm p})}{dE_{\pi}}
\end{equation}
 where $n_{\rm target}$ stands for $\rho/m_{\rm p}$, and  
 $E_{\rm p,\pi}$ and  $p_{\rm p,\pi}$ denote the
 energy and momentum of protons or pions, respectively.
 Full descriptions of the integral limit and $\sigma_{\rm pp\rightarrow\pi}$ 
 are given in \citet{naitotakahara94}.
 The contributions of the inverse Compton (IC) and Bremsstrahlung are 
 calculated from $j_{\rm e,p}(E_{\rm e,p})$ as 
\begin{equation}
\label{eq:gamma_flux-electrons}
 F_{\gamma}^{\rm IC, Brem}(E_{\gamma}) = \frac{1}{D^2}\int{n_{\rm target} dV}
 \int^{E^{\rm max}_{\rm e}}_{m_{\rm e} c^2} dE_{\rm e} j_{\rm e}(E_{\rm e}) 
 \frac{d\sigma}{dE_{\gamma}}, 
\end{equation}
 where $n_{\rm target} = n_{\rm photon}$ 
 and $\frac{d\sigma}{dE_{\gamma}}$ is a cross section which
 includes the Klein-Nishina effect for inverse Compton emission, and 
 $n_{\rm target}$ of $\rho/m_{\rm p}$ 
 and the cross section $\frac{d\sigma}{dE_{\gamma}}$ 
 of electron-proton and electron-electron interaction 
 are used for Bremsstrahlung emission \citep{gaisser98,sturner97}. 
 For $n_{\rm photon}$, we adopt 2.7~K CMB and $T_{\rm eff} =$ 27,000~K 
 black body radiation from the Be star.
 In the spatial integration, we assume that the accelerated particles extend 
 into the Be star outflow downstream of the shock.
 The contributions of different emission mechanisms are calculated 
 with $x_{\rm disc} =$1500 for the phase of periastron 
 and their differential energy spectra are 
 shown in Fig.\,\ref{fig:spectrum_comparison}
 (the disc and the pulsar wind are assumed to interact at
 periastron in the calculation).
 The total gamma-ray flux is deduced as 
 $F_{\gamma}(E_{\gamma}) =  F_{\gamma}^{\rm Brem}(E_{\gamma}) +
 F_{\gamma}^{\rm IC}(E_{\gamma}) +  F_{\gamma}^{\rm pp}(E_{\gamma})$ and 
 the dominant contribution is of $F_{\gamma}^{\rm pp}(E_{\gamma})$.
 The inverse Compton flux in the sub-TeV energy region, expected from 
 the pulsar-wind side\,\citep{shibazaki02}, is comparable to 
 $F_{\gamma}^{\rm IC}$ from the Be star outflows, except that the former has 
 a break $\sim$400 GeV due to the stronger magnetic field in the 
 pulsar wind side.
 After the spatial integration, the total flux is approximately 
 expressed as 
\begin{equation}
\label{eq:gamma_flux-propt}
 F_{\gamma}(E_{\gamma}, x) 
 \propto x^2 \frac{1}{n-1} \frac{1}{R_{\rm shock}(x,n,v_0)}~~.
\end{equation}
 $R_{\rm shock}(x,n,v_0)$ for the same orbital phase does
 not vary much within the parameter range discussed in the following. 

 The adopted model parameters are summarized in 
 Table\,\ref{table:model_parameters}.
 Now we discuss the possible ranges of two parameters,
 $x$ and the density profile index $n$. 
 For the polar component, the value $x_{\rm polar}$ is set to be 
 proportional to the disc component, $x_{\rm disc}$. 
 The factor is estimated using the following two equations, 
 $\dot{M}_{\rm polar}
 =4 \pi R_*^2~\rho_{\rm polar, 0}~v_{\rm polar, 0}~(1-\sin \theta_{\rm disc})$ and  
 $\dot{M}_{\rm disc}
 =4 \pi R_*^2~\rho_{\rm disc, 0}~v_{\rm disc, 0}~\sin \theta_{\rm disc}$, 
 where the surface density, initial velocity, 
 and mass loss rate of the disc [polar wind] flow 
 are denoted as $\rho_{\rm disc[polar], 0}, v_{\rm disc[polar], 0},$ and 
 $\dot{M}_{\rm disc[polar]}$, respectively,
 and $\theta_{\rm disc}$ denotes the  opening angle of disc outflow. 
 The ratio of two $x$ parameters is 
\begin{equation}
 \frac{x_{\rm polar}}{x_{\rm disc}} = 
 3.49 \times 10^{-1}
 \frac{\dot{M}_{\rm polar}}{\dot{M}_{\rm disc}}, 
\end{equation}
 assuming $\theta_{\rm disc}$ = 15\arcdeg \citep{johnston96}.
 From the observed intensities of the UV line (due to the polar wind) and 
 of the IR radiation (from the disc),  
 \citet{lamers87} deduce  
 the mass loss ratio of the two flow components as 
 $\frac{\dot{M}_{\rm polar}}{\dot{M}_{\rm disc}}$ of 10$^{-1}$--10$^{-4}$. 
 We take $x_{\rm polar}$ of 10$^{-1} \times x_{\rm disc}$ 
 as a rather optimistic value.
 The thick disc-like flow is an effective site for the production of 
 gamma-ray emission, and makes the dominant contribution 
 to the total intensity.

 For $x_{\rm disc}$, early studies \citep{waters86, waters88, dougherty94} 
 have estimated possible ranges of 
 ${\rm 10}^5 < v_{\rm disc, 0} < {\rm 10}^7 {\rm cm s}^{-1}$  and 
 ${\rm 10}^{-13} < \rho_{\rm disc, 0} < {\rm 10}^{-9} {\rm g cm}^{-3}$. 
 Thus we investigate 500 $\le x_{\rm disc} \le$ 5000 in this model analysis.

 In Eq.\,\ref{eq:p_be},  $n = $2 corresponds to a constant speed and 
 $n = $4 corresponds to a constant ram pressure. 
 We can approximate the polar wind with $n_{\rm polar} =$2, 
 since the polar wind 
 is generally thought to reach a terminal speed within a few 
 stellar radii. In contrast, the disc density falls rapidly with radius  
 as the rotating material gradually accelerates outward.
 \citet{ball99} describes the disc of SS\,2883 with 
 $n_{\rm disc} \lesssim$ 4, while  
 \citet{waters88} gives 2 $< n_{\rm disc} <$ 3.25 from 
 a general consideration of Be star discs.  
 We adopt $n_{\rm disc} = $2.5 here. 
 Changing $n_{\rm disc}$  to 4 and keeping other parameters fixed 
 reduces the emission by a factor of about 2
 (Eq. \,\ref{eq:gamma_flux-propt}).  

 There are no fixed limits for the orbital phases in which 
 the Be star disc outflow interacts with the pulsar wind. 
 We consider three possibilities in calculating the light curves;  
(i) aligned disc to the orbital plane and interaction 
  throughout the orbit, 
(ii) mis-aligned disc and interaction 
 in the $\sim$200-day period around periastron ($\tau$), 
 during which the radio emission is depolarized, or 
(iii) mis-aligned disc and interaction in two short periods, 
[($\tau - $18~d) $\sim$ ($\tau - $8~d)] and [($\tau +$12~d) 
 $\sim$ ($\tau +$22~d)],  as discussed in 
 \,\citet{connors02}.
 Eq.\,\ref{eq:gamma_flux-propt} with 
 $x_{\rm polar}$ of 10$^{-1} \times x_{\rm disc}$, suggests that the 
 contribution from the polar-wind--pulsar-wind interaction is a factor 
 1.5$\times {\rm 10}^{-2}$ of that from the disc--pulsar-wind interaction.
 The polar wind is generally assumed to 
 interact with the pulsar wind at all orbital phases.
 When the disc and pulsar-wind interaction diminishes. 
 the estimated intensity from the system is only of 
 the polar-wind contribution, and is reduced by a factor of 
 $\sim {\rm 10}^{-2}$.
  We take account of (i), implying the maximum effect of the disc-pulsar 
 wind interaction,  though the disc material becomes dilute at larger
 distances
 (Eq.\,\ref{eq:p_be}).
 In (ii) and (iii), we consider emissions from the pulsar-wind bubble 
 formed in the disc flow, after the pulsar leaves the disc\,\citep{connors02}.
 The bubble moves at the velocity $v_{\rm bubble}$ in the outflow 
 and shock acceleration of particles in the flow proceeds
 in the contact discontinuity between the bubble and the outflow material.
 Emissions from the moving bubble are calculated along its trace
 referring to the material and momentum density profiles of the flow,  
 by replacing  $R_{\rm shock}(x,n,v_0)$ with 
 $R_{\rm shock}(t=t_0, x,n,v_0) + v_{\rm bubble}(t-t_0)$  in 
 Eq.\,\ref{eq:gamma_flux-propt}, where $t_0$ denotes the time when the pulsar 
 moves out of the disc flow.
 We assume an initial value of $v_{\rm bubble} =$ 100${\rm km}{\rm s}^{-1}$ 
 which is larger than the value used in \citet{connors02}, 
 15${\rm km}{\rm s}^{-1}$, but is similar to the model in 
 \citet{paredes91} as well as to the typical velocity of the disc flow.
 The adiabatic expansion, which is mainly important for synchrotron emission,
 does not affect much the emission mechanism mentioned here.
 The rise time of bubble emission is assumed to be 
 $\sim$ 1 day.

\subsection{Comparison with the Results}
\begin{figure}[ht]
\plotone{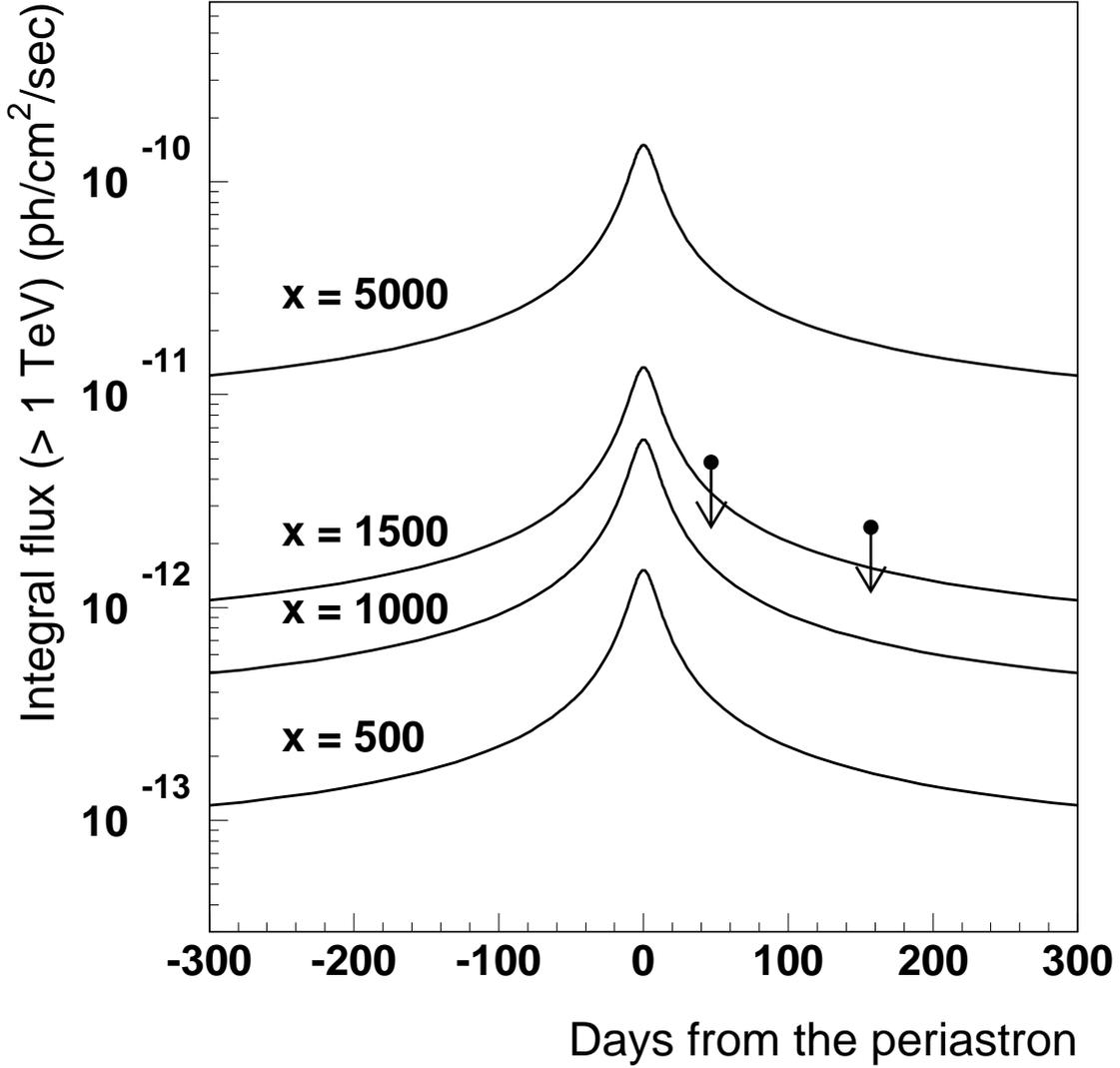}
 \caption{
 Upper limits on the integrated gamma-ray flux 
 are compared with the model calculated curves (E $\ge$ 1\,TeV) 
 as a function of days from periastron epoch.
 The energy thresholds of our results have been scaled to 1\,TeV 
 assuming a $E^{-2.0}$ spectrum for comparison with the predictions.
 The light curves of the combined flux from the Be star 
 outflows are calculated with the model assumption (i) (see the text)  
 and $x_{\rm disc} =$ 500, 1000, 1500,  and 5000.
}  
\label{fig:model_comparison_x}
\end{figure}

\begin{figure}[ht]
\plotone{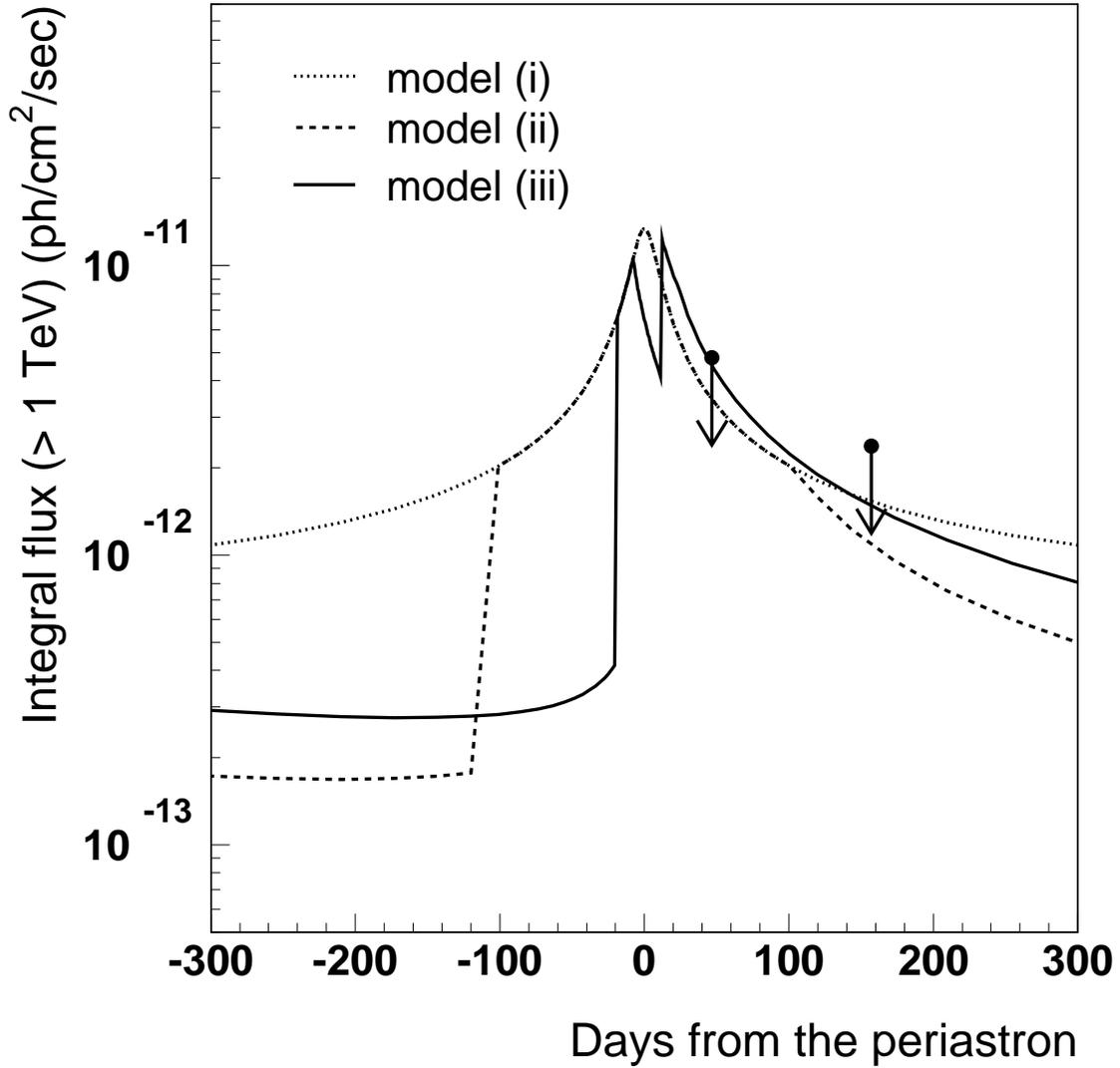}
 \caption{
 Our upper limits and the model calculated light curves (E $\ge$ 1\,TeV) 
 are compared as in Fig.\,\protect\ref{fig:model_comparison_x}.
 The light curves with $x_{\rm disc}$ of 1500 are 
 calculated with the model assumptions (i)--(iii);   
 where (ii) and (iii)  the disc and the pulsar wind are assumed to interact 
 only for the limited period(s).
}  
\label{fig:model_comparison_interact}
\end{figure}

 The observational upper limits are compared with light curves 
 calculated from the model.
 The energy thresholds of our results have been scaled to 1\,TeV 
 assuming a $E^{-2.0}$ spectrum. 
 The spectra calculated with the model assumption (i) in 
 disc-pulsar wind interaction, 
 are integrated  (E $\ge$ 1\,TeV)  for four different mass outflow 
 parameters, $x_{\rm disc} =$ 500, 1000, 1500, and 5000 
 (Fig.\,\ref{fig:model_comparison_x}).
 The outflow parameter is constrained by our results to 
 $x_{\rm disc} \le $1500. 
 The light curves with the different 
 model assumptions (i)--(iii) for the fixed mass outflow parameter  
 $x_{\rm disc}$ of 1500 are shown in  
 Fig.\,\ref{fig:model_comparison_interact}.
 As discussed in the previous subsection, 
 the light curve is reduced by a factor of $\sim$ 10$^{-2}$
 outside the assumed disc--pulsar-wind interaction period
 since the polar-wind becomes the only counterpart of 
 the pulsar-wind. In addition, 
 contribution from the wind-bubble formed in the disc--pulsar-wind 
 interaction, remains while the bubble is moving in the disc.
 Thus, for the model assumption (iii) where the disc and the pulsar-wind 
 interact twice in the orbit, the emission peak after periastron consists 
 of the ``second'' disc--pulsar-wind interaction, 
 of the disc-wind-bubble interaction where the bubble is the outcome 
 of the ``first'' interaction, 
 and of the polar-wind--pulsar-wind interaction.
 For all three assumptions the constraint from the observations, mainly from 
	$Obs.\,A$, is similar.
 With this relatively small outflow pressure, the Be star wind may not be 
 able to overwhelm the pulsar wind pressure to produce accretion onto
 the pulsar,
 as has been suggested by the X-ray observations.

 Besides our emission models based on the Be-star outflows, 
 we discuss the light curve shown in Fig.\,5 of \citet{ballkirk00} 
 as the optimum case for TeV emission from the pulsar wind side, 
 using rather ideal model of the inverse Compton scattering on 
 the un-shocked pulsar wind with a wind Lorentz factor of 10$^7$.
 Our upper limits are modified into units of integral energy flux 
 using approximated spectral indices in Fig.\,4 of  \,\citet{ballkirk00},  
 but the obtained limit
 of $\sim$1$\times {\rm 10}^{-5} {\rm MeV} {\rm cm}^{-2}{\rm s}^{-1}$, 
 does not strongly constrain the model since 
 the light curve quickly declines from $\sim$5 
 at the periastron epoch 
 to 0.2 in units of ${\rm 10}^{-5} {\rm MeV} {\rm cm}^{-2}{\rm s}^{-1}$. 
 The integrated flux greater than 1~TeV is obtained from 
 another model calculation of pulsar wind emission using the 
 spectra in Fig.\,7 of \citet{shibazaki02}. 
 They argue for the dominance of synchrotron cooling in 
 the energy loss of the pulsar wind electrons.
 Assuming the distance of 1.5\,kpc, the  predicted flux of 
 ${\rm 10}^{-14} {\rm cm}^{-2} {\rm s}^{-1}$  is 
 about two orders of magnitude smaller than our limit.

\begin{figure}[ht]
\plottwo{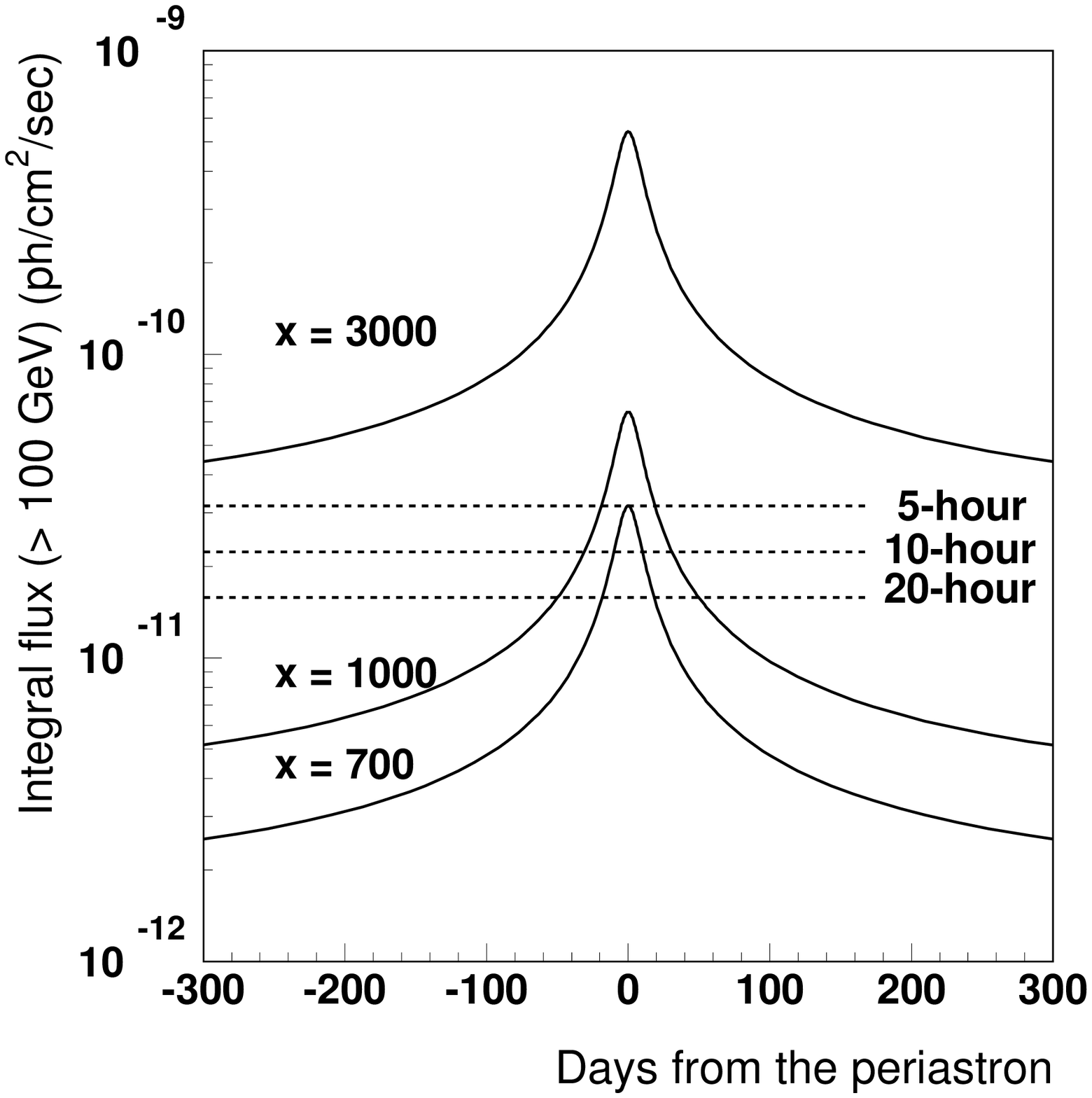}{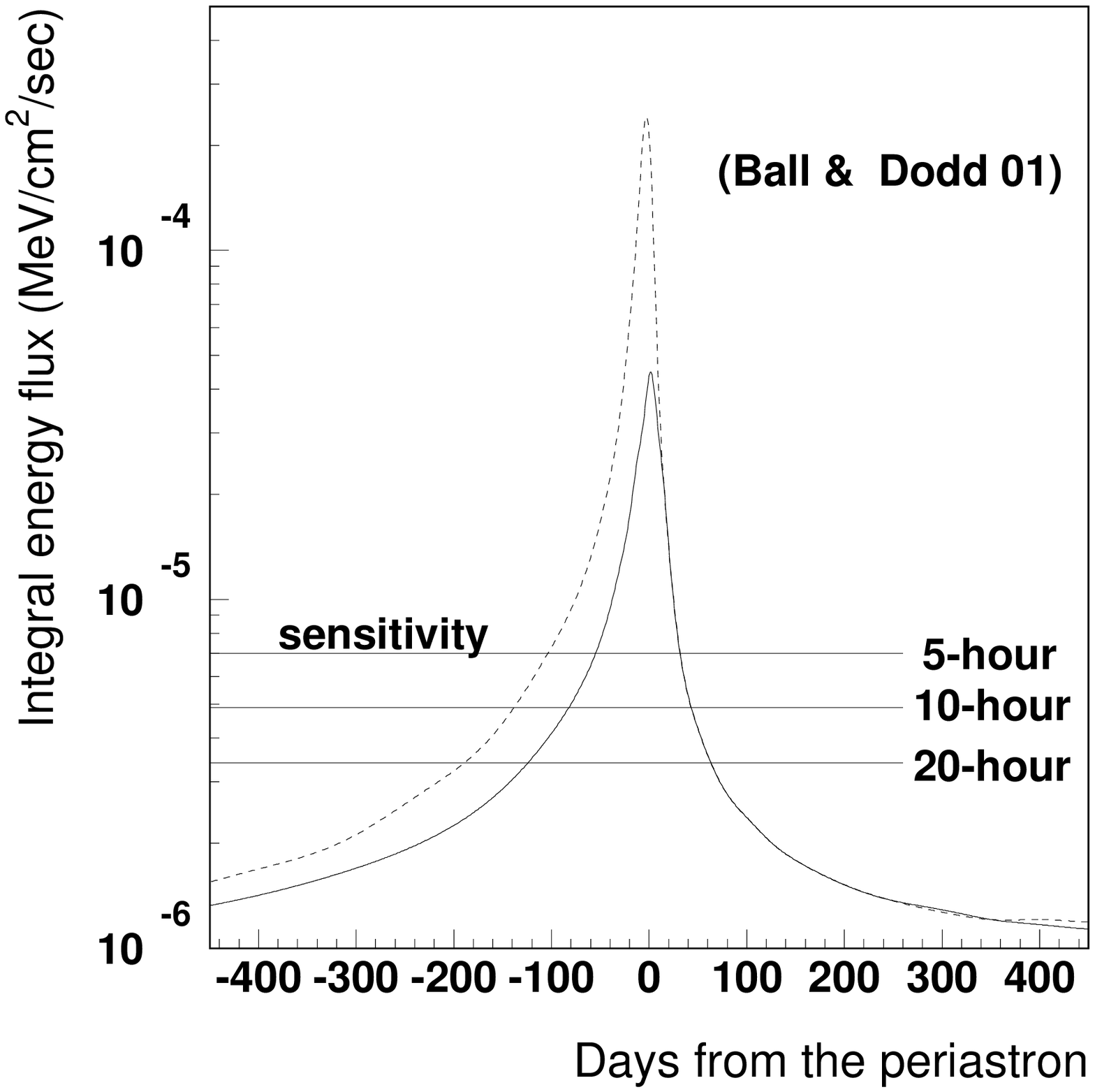}
 \caption{
 The expected sensitivities of the state-of-art Cherenkov telescopes 
 are compared with the models.  
 ($left):$  Corresponding to the improved sensitivity,  
 light curves of our model are re-calculated changing the integral limits, 
 from 1~TeV to 100\,GeV. The sensitivity levels of 20, 10, 
 and 5-hour  observations are drawn in dotted lines.
 $x_{\rm disc}$ of 700, 1000 and 3000 are tried.
 ($right):$  The estimated light curves of 100~GeV emissions 
 from the pulsar wind are taken from Fig.\,5 of \protect\citet{balldodd01}.
 Calculated results for the terminated and un-terminated pulsar wind models
 are shown in solid and dotted lines, respectively.
 The unit of the expected sensitivities (20, 10, and 5-hour) are modified in 
 ${\rm MeV} {\rm cm}^{-2}{\rm s}^{-1}$, using the approximated spectral 
 shape of Fig.\,4 of \protect\citet{balldodd01}.
}  
\label{fig:future_comparison}
\end{figure}

 Recently, new projects of ground based Cherenkov telescopes have begun 
 operations\,\citep{review-latest}. 
 With the improved sensitivity and the lower energy threshold, they will 
 offer a better opportunity to observe the PSR\,B1259$-$63 binary system 
 in the high-energy band.
 For projects such as CANGAROO-III or H.E.S.S., 
 located in the southern 
 hemisphere, a 50-hour observation of the binary system gives 
 a typical sensitivity of 
 $\sim {\rm 10}^{-11} {\rm cm}^{-2} {\rm s}^{-1}$ 
 with the energy threshold of $\sim$100\,GeV\,\citep{hess-performance}.
 The calculated spectra of our models are integrated again, for the 
 energy greater than 100\,GeV, for comparison with this sensitivity.
 In Fig.\,\ref{fig:future_comparison}, the sensitivity levels 
 of 20-hour, 10-hour and 5-hour (statistically scaled) observations, 
 respectively, are drawn over the calculated light curves.
 A day-scale light curve might be detectable for the model with 
 $x_{\rm disc} \ge \sim$700 along the periastron passage.
 \citet{balldodd01} estimate the $\sim$100~GeV emission from the 
 pulsar wind with a Lorentz factor of 10$^6$, and their light curves 
 are compared with these expected sensitivities 
 after modification of the unit into the integral 
 energy flux (${\rm MeV} {\rm cm}^{-2}{\rm s}^{-1}$), assuming 
 the spectral shape (Fig.\,4 of \,\citet{balldodd01}).
 The light curves in Fig.\,\ref{fig:future_comparison} $(right)$ 
 are taken from Fig.\,5 of \,\citet{balldodd01} showing terminated 
 (solid line)
 and un-terminated (dashed line) shock models in the pulsar wind emissions. 
 Both model predictions  are comparable with the detectable flux, 
 at least, around the periastron epoch.
 From \citet{shibazaki02}, integrations $E \ge$ 100~GeV are performed 
 resulting in fluxes of $\sim$4$\times {\rm 10}^{-12} 
 {\rm cm}^{-2} {\rm s}^{-1}$ at periastron and
 $\sim$1$\times {\rm 10}^{-12} {\rm cm}^{-2} {\rm s}^{-1}$ at apastron, which 
 is still below the improved sensitivity of 
 ground based detectors.

\section{Summary}
 The PSR\,B1259$-$63/SS2883 binary system has been observed at TeV energies
 using the CANGAROO-II 10-m telescope. 
 The observations were performed at two different orbital phases 
 following the October 2000 periastron.
 Upper limits on the integrated TeV gamma-ray flux are obtained.
 A new model for gamma-ray emission from the Be star outflow 
 has been introduced, and contributions from 
 Bremsstrahlung, inverse Compton scattering, and proton-proton interactions, 
 are calculated, with possible variations in parameters considered.
 The light curves are calculated with different assumptions on 
 the disc-pulsar wind interaction.
 The estimated light curves are discussed and compared with our 
 observational results to 
 constrain the disc-like outflow 
 density $\rho_{0,-12} = \rho_{0}/{\rm 10}^{-12}$ g cm$^{-3}$ 
 and its flow speed $v_{0, 6} = v_{0}/10^{6}$ cm s$^{-1}$ 
 by $x_{\rm disc} = \rho_{0,-12}~v_{0, 6} \le$ 1500. 
 The next periastron will occur in March 2004 when 
 the condition will favorable for small zenith angle observations and
 hence low energy thresholds for ground based Cherenkov telescopes.
 Further observations of PSR\,B1259$-$63 system 
 during the periastron passage (including the weeks before and after 
 the periastron, respectively) are encouraged to 
 provide valuable information.

\acknowledgments
 The authors greatly thank Dr.\, L.~Ball for introducing his
 model calculations to us and for promoting the first idea of 
 these observations. 
 Dr.\,T.~Terasawa gave useful comments in the early discussion of 
 the model. 
 The research was supported by a Grant-in-Aid for 
 Scientific Research of the Ministry of Education, Culture, 
 Science, Sports and Technology of Japan and 
 by the Australian Research Council.
 AA, SH, JK, LK, KO, KS, and KT were supported by Research Fellowship 
 and Postdoctoral Fellowships of Japan Society of Promotion of Science.

\clearpage

\clearpage

\clearpage

\clearpage

\clearpage
\clearpage

\clearpage

\clearpage

\clearpage

\clearpage
\begin{table}
\begin{center}
\caption{CANGAROO-II observations of the PSR B1259$-$63/SS2883 
 binary.  The days after the October 2000 periastron are calculated with 
 the average MJD of the observation periods.
 \label{table:obs_summary}
}
\begin{tabular}{lcccc}
\\
\tableline
\tableline
Observation & \multicolumn{2}{c}{Epoch} &  True Anomaly\tablenotemark{a} & Separation\tablenotemark{b} \\
            & (UT) & (MJD)              &  (degree) &     (10$^{13}$cm)  \\
\tableline
Obs.\,A & & & \\
	& 2000/12/01 & 51879.72--51879.76 &  & \\ 
	& 2000/12/03 & 51881.69--51881.74 &  & \\ 
	& 2000/12/04 & 51882.71--51882.76 &  & \\ 
        & & & \\
        & \multicolumn{2}{c}{ave. 51881.4, periastron $+$ 47 days} & 125 
                     & 3.6 \\
\tableline
Obs.\,B & & & \\
	& 2001/03/19 & 51987.60--51987.68 &  & \\ 
	& 2001/03/21 & 51989.62--51989.75 &  & \\ 
	& 2001/03/22 & 51990.62--51990.79 &  & \\ 
	& 2001/03/24 & 51992.59--51992.73 &  & \\ 
	& 2001/03/25 & 51993.54--51993.70 &  & \\ 
	& 2001/03/26 & 51994.53--51994.60 &  & \\ 
        & & & \\
        & \multicolumn{2}{c}{ave. 51991.5, periastron $+$ 157 days} & 153 
		& 8.1 \\
\tableline
\end{tabular}\\
\tablenotetext{a}{The true anomaly is zero at the epoch of periastron}
\tablenotetext{b}{The binary separation at periastron is assumed to be 
 0.97$\times {\rm 10}^{13}$cm}
\end{center}
\end{table}

\clearpage
\begin{table}
\begin{center}
\caption{Summary of the observations and the data reduction. 
 In the effective observation time, 
 data taken in poor weather conditions have been rejected 
 and the dead-time due to the data acquisition process has been corrected for.
 The cuts applied to the camera images
 are described in the text.
 The average zenith angle $\theta_{zen}$ is calculated 
 for the whole observation set. The energy threshold $E_{th}$ has been 
 deduced assuming a spectral index of $-$2.5 and   
 is identical in the ON- and OFF-source data.
\label{table:data_summary}
}
\begin{tabular}{lccccccc}
\\
\tableline
\tableline
Data & $\theta_{zen}$  & $E_{th}$ & \multicolumn{2}{c}{Time} &  
                \multicolumn{3}{c}{Number of events} \\ 
   & & &  Real & Effective &  Recorded  & Noise  &  Image  \\
   & & &       &           &            &  Reduction &  Selected \\
   & (degree) & (TeV) & (min) & (min) &   & & \\
\tableline
 $Obs.\,A$  & & & & & & & \\
    ON:  & 58.9 & 3.6  & 202   &  196 & 1.17E5 & 6.62E3 & 1.85E3 \\ 
    OFF: & 59.2 &      & 193   &  160 & 4.29E5 & 5.86E3 & 1.62E3 \\
\tableline
$Obs.\,B$  &  & & & & & & \\
    ON:  & 34.0 & 0.78 & 1078  &  623 & 4.43E6 & 7.06E4 & 1.81E4 \\
    OFF: & 34.3 &      & 1001  &  645 & 2.63E6 & 7.21E4 & 1.87E4 \\
\tableline
\end{tabular}
\end{center}
\end{table}

\clearpage
\begin{table}[ht]
\begin{center}
\caption{Adopted parameters in the model calculation 
	of gamma-ray emissivity. See the text for the detailed definitions. 
}
\label{table:model_parameters}
\begin{tabular}{lr}
\\
\tableline
\tableline
Parameters & Adopted Values \\
\tableline
{\it Pulsar wind :} &  \\
Fraction to the equatorial plane, f$_{\rm pw}$  &  0.1 \\
\tableline
{\it Be star :} &  \\
Radius, $R_*$ & 6$R_{\sun} = {\rm 4.17} \times {\rm 10}^{11}{\rm cm}$\tablenotemark{a}\\
Distance, $D$ & 1.5\,kpc\tablenotemark{a}\\
Opening angle of the disc outflow, $\theta_{\rm disc}$ & 15 degree \\
Power law index of density profile of the disc, $n_{\rm disc}$ & 2.5 \\
Power law index of density profile of the polar wind, $n_{\rm polar}$ & 2  \\
Efficiency of the acceleration, $f_{\rm acc}$ & 0.1 (proton) \\
 &  0.001 (electron)    \\
Power law index of the proton/electron energy flux, $\alpha$ &  $-$2.0 \\
Maximum energy of the accelerated particles, 
$E^{\rm max}_{\rm e,p}$ & 10$^{15}$~eV\\
 Outflow parameter of the disc, $x_{\rm disc}$ 
 & 500 -- 5000 \\
Outflow parameter of the polar wind,  $x_{\rm polar}$ & 
  $10^{-1} \times x_{\rm disc}$\\
\tableline
\end{tabular}
\tablenotetext{a}{Taken from \protect\citet{johnston94}}
\end{center}
\end{table}
\end{document}